\documentclass[10pt,journal,twocolumn]{IEEEtran}
\usepackage{graphicx}
\usepackage{dcolumn}
\usepackage{amsmath,amsfonts}
\usepackage{bm}
\usepackage{multirow}
\usepackage{slashbox}
\usepackage{tabularx}
\usepackage{tikz}
\begin{document}
\title{Cross-scale Attention Model for Acoustic Event Classification}
\author{Xugang Lu$^{1*}$, Peng Shen$^{1}$, Sheng Li$^{1}$, Yu Tsao$^{2}$, Hisashi Kawai$^1$
\thanks{1. National Institute of Information and Communications
Technology, Japan. Email: xugang.lu@nict.go.jp
}
\thanks{2. Research Center for Information Technology Innovation, Academic
Sinica, Taiwan} }

\IEEEcompsoctitleabstractindextext{%
\begin{abstract}
\small{}
A major advantage of a deep convolutional neural network (CNN) is that the focused receptive field size is increased by stacking multiple convolutional layers. Accordingly, the model can explore the long-range dependency of features from the top layers. However, a potential limitation of the network is that the discriminative features from the bottom layers (which can model the short-range dependency) are smoothed out in the final representation. This limitation is especially evident in the acoustic event classification (AEC) task, where both short- and long-duration events are involved in an audio clip and needed to be classified. In this paper, we propose a cross-scale attention (CSA) model, which explicitly integrates features from different scales to form the final representation. Moreover, we propose the adoption of the attention mechanism to specify the weights of local and global features based on the spatial and temporal characteristics of acoustic events. Using mathematic formulations, we further reveal that the proposed CSA model can be regarded as a weighted residual CNN (ResCNN) model when the ResCNN is used as a backbone model. We tested the proposed model on two AEC datasets: one is an urban AEC task, and the other is an AEC task in smart car environments. Experimental results show that the proposed CSA model can effectively improve the performance of current state-of-the-art deep learning algorithms. 
\end{abstract}
}

\maketitle

\IEEEdisplaynotcompsoctitleabstractindextext

\IEEEpeerreviewmaketitle

\section{Introduction}
\label{sec-I}
The acoustic event classification (AEC) task is to annotate given audio streams with their semantic categories. It is an important step for audio content analyses and audio information retrieval \cite{Giannoulis et al 2013, Heittola et al 2013, Zhuang et al 2010}. In contrast to automatic speech recognition (ASR), where acoustic and language models can be defined based on a clear hierarchical structure of speech (e.g., phones, syllables, and words), acoustic events do not have phone- or word-like units to connect low-level acoustic features to high-level semantic labels in modeling. Therefore, successful model pipelines used in ASR are unsuitable in AEC. Most conventional AEC algorithms are based on a two-step process framework: feature extraction and classification modeling. In feature extraction, the bag-of-frames (BoF) approach is often applied \cite{Plinge2014,Grzeszick2017}. Based on the BoF approach, acoustic events are represented as a histogram distribution of basic frame features. To derive long temporal information for event representation, the bag-of-acoustic-word model  has been proposed, where acoustic events are represented as a histogram distribution of basic temporal-frequency spectral patches or acoustic words \cite{Lu2014,Cotton et al 2010}. Considering these representations, Gaussian mixture models and support vector machines are often adopted to form classifiers \cite{Temko 2008,Huang2013}. To take the temporal correlation between acoustic frames and words into consideration, the hidden Markov model has also been adopted in modeling \cite{Zieger 2008}. For these conventional AEC algorithms, the feature extraction and classifier modelling processes are designed independently and not optimized in a joint learning fashion.

With the successful applications of supervised deep learning (SDL) frameworks in processing and recognizing image and speech signals, they have also been applied in the AEC task. The advantage of the SDL frameworks is that they can automatically learn discriminative features and classifiers in a joint learning fashion. Numerous models with various types of network architectures have been proposed based on the SDL frameworks. Among them, the convolutional neural network (CNN) model can explore temporal- and/or  frequency-shift invariant features of AEC \cite{Salamon2017,Piczak2015,Gorin16}.  The recurrent neural network (RNN) model can extract long temporal-context information in feature representation for classification. With long short-term memory (LSTM) units \cite{Hochreiter97}  or gated recurrent units (GRUs) \cite{Cho14}, the RNN can be efficiently trained for AEC. Models that combine the advantages of the CNN and RNN have also been proposed, e.g., convolutional recurrent neural network (CRNN) model, where the CNN is used to explore invariant features, and the RNN is used to model the temporal structure in a classification \cite{Emre17,ChoiICASSP17}.

Among the deep-learning-based models, the CNN has become an indispensable building block for designing state-of-the-art systems. In a typical pipeline for AEC based on a deep learning framework, multiple layers of CNNs are used for feature extraction. By stacking multiple CNN layers, hierarchical multi-scale features are explored. Features extracted from the top layers encode long-range feature dependency because  top-layer convolutions correspond to large scales of receptive fields. Therefore, features in the final representation layer encode the global and abstract characteristics of acoustic events. Although feature representations from the top layers are robust or invariant to various local variations, the local representations from the bottom layers, which are important in discriminating different acoustic events, may be smoothed out in the final representations. However, an ACE task generally contains both short- and long-duration events that are involved in audio clips, and thus discriminative features from local regions should be efficiently propagated to the final representation for classification. Accordingly, several classification models that integrate multi-scale features or multi-scale decision  scores have been proposed \cite{LeeDCASE2017,GuoDCASE2018}. In most of these works, rather than utilizing the hierarchical-scale features extracted from stacked CNN layers, they used multi-scale spectral patches as the input to stacked CNN layers to extract multi-scale features or multi-scale scores. Although improved performance can be achieved, repeatedly passing multi-scale spectral patches in the model may cause redundant calculations, which increase the computational cost.

Besides taking advantage of integrating multi-scale hierarchical features for classification, the attention mechanism has been popularly integrated in deep CNN (DCNN) models for the AEC task. Notable examples include the attention- and localization-based deep models \cite{KongICASSP17, KongIEEE09, XuIS17, XuNo117}  and attention pooling algorithm \cite{Ilse2018, LuIS2018, Kumar2016}. Generally, these studies were inspired by the success of the attention mechanism in machine translation and image processing \cite{Hahdanau14, Luong15, Chan16}. Most of these algorithms consider either temporal or spatial attention to localize the important time or time–frequency regions and conduct pooling on features along the temporal or frequency dimensions. The final feature representation used to perform classification is based on the features from the last convolutional block, which has a large receptive field. Although the AEC performance has been notably enhanced by the attention mechanism concept, the several advantages of feature maps in DCNN have not been fully utilized. For example, the features are actually explored and represented in spatial (time–frequency), multi-channel (number of feature maps), and hierarchical structures (multi-scales), and these features represent different aspects of acoustic events. We argue that a model that can dynamically determine the suitable scale for a specific acoustic event may further improve the overall classification performance.

In summary, the semantic meanings of sound events are attributed to the different levels of abstractions; that is, the discriminative features of sound events are distributed in multiple scales with local and global dependencies. In this paper, we propose a cross-scale attention (CSA) model for the AEC task. The CSA model dynamically weights features in different scales based on their importance and aggregates cross-scale features for AEC. Although algorithms using multi-scale features and attentions have been proposed \cite{Chou2018, Yu2018}, where the multi-scale features are concatenated or used as inputs to another network for discriminative feature extraction, in our model, the small-scale features are progressively and adaptively propagated to the top layers for discriminative feature extraction. As regards the functional effect, our proposed CSA model is quite similar to the residual CNN (ResCNN), which progressively passes small-scale features via skip connections from the bottom to top layers \cite{HeCVPR2016, HeECCV2016}. However, the ResCNN uses fixed and implicit scale feature weighting, and its initial motivation with skip connections is to deal with the vanishing gradient problem in training deep networks \cite{HeCVPR2016, HeECCV2016}. In our CSA model, we explicitly use adaptive scale weighting for feature propagation from the bottom to top layers. In this way, the final feature representation encodes local and global discriminative features with a wide range of scales, which are expected to improve the performance of the AEC task. Our contributions are summarized as follows:

(1) We explicitly formulated the importance of features in consecutive scales from stacked CNN layers with an attention model, which facilities the DCNN to adaptively propagate small-scale features to the top layers. The final representation encodes discriminative features from a wide range of scales with consideration of their importance for the target acoustic events.

(2) We mathematically revealed the relationship between our CSA model and ResCNN models. The ResCNN model can be regarded as a special case of implicit fixed-scale feature propagation in feature extraction. In addition, based on the design of the scale attention function, temporal and spatial context dependencies can be taken into consideration. In this sense, the proposed CSA model can be regarded as a unified attention model by taking temporal and/or  spatial information in attention models.

(3) We proposed a system for AEC based on the CSA model and evaluated the performance of the system by exploring several factors, which may affect the performance.

The remainder of the paper is organized as follows. Section \ref{sec-II} introduces the background and fundamental theory of the proposed CSA model and explains its connection to the ResCNN. Section \ref{sec-III} describes the implementation details of the CSA model for the AEC task. Section \ref{sec-IV} presents the AEC experiments and results based on the proposed framework by analyzing the contribution of the CSA model in detail. Section \ref{sec-V} presents the discussion of the results and conclusion of the study.

\section{Background and theory of the CSA model}
\label{sec-II}
Because the discriminative features of acoustic events are encoded in multi-scale patterns, we need to examine two problems: (1) how to extract multi-scale features from acoustic signals and (2) how to weigh and integrate these multi-scale features based on their importance in class discrimination. In this section, we first explain the multi-scale feature extraction in the DCNN and then introduce the proposed CSA model and the design of the AEC framework based on the CSA model.

\subsection{Multi-scale feature representations in a DCNN}
\label{sec-II-I}
\begin{figure}[tb]
\centering
\includegraphics[width=7cm, height=4cm]{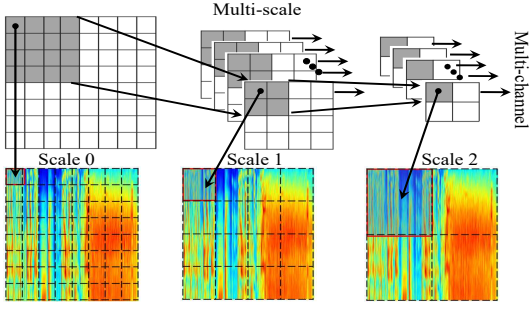}
\caption{Multi-scale feature representations in a DCNN: convolution in top layers can ``look" more wider with large scale of receptive fields than in bottom layers.}
\label{fig1}
\end{figure}

Intuitively, the multi-scale patterns of acoustic events are distributed in various types of temporal-frequency patterns. Therefore, we define the temporal-frequency spectral patches as instances of acoustic events. Based on this definition, acoustic events can be described as bags of spectral patches. In addition, to consider different sizes of instances in modelling, the bags can be composed of multi-scale spectral patches. With this concept, we can explain the function of the DCNN-based acoustic feature extraction as a process of multi-scale feature detection. To explain our proposed architecture, we first describe the formulations of the DCNN. The DCNN is composed of multiple processing blocks, and each block consists of an affine transform (convolution), nonlinear activation, and feature pooling, which are represented as follows:
\begin{equation}
\begin{array}{l}
 {\bf V}^l  = {\rm Conv}\left( {{\bf X}^{l - 1} } \right) \\
 {\bf O}^l  = f_{{\rm ReLU}} \left( {{\bf V}^l } \right) \\
 {\bf X}^l  = f_{{\rm pool\_max}} \left( {{\bf O}^l } \right), \\
 \end{array}
\label{eq1}
\end{equation}
where $l$ is layer index, ${\rm Conv}\left(  \circ  \right)$ is a linear convolution process with 2-dimension (2D) kernels (convolution along both the time and frequency dimensions), $f_{{\rm ReLU}}  \left(  \circ  \right)$ is a neural activation function (rectified linear unit (ReLU) is used in this study), and $f_{{\rm pool\_max}} \left(  \circ  \right)$ is a max-pooling (MP) operation. The output of the $l$-th layer ${\bf X}^l  \in {\mathbb R}^{k_l  \times F_l  \times T_l }$ is a 3-dimension (3D) tensor, where $k_l$, $F_l$, $T_l$ are the number of CNN filters (or filter channels), dimensions of height and width of feature maps, respectively. The hierarchical feature structure corresponding to multi-scale spectral patches is encoded in the tensors, as shown in Fig. \ref{fig1}.
In the figure, there are two layers of convolutional blocks (function modules are defined as in Eq. \ref{eq1}). The first input layer ``Scale 0" corresponds to the raw spectral space. The convolutional filters in each layer can be regarded as spectral patch detectors. Each ``pixel" in the output space of convolutional blocks is a representation corresponding to the different scales of a spectral patch (with different sizes of receptive fields). For example, as shown in Fig. \ref{fig1}, each pixel in ``Scale 1" or ``Scale 2" is a representation of a $2*2$ or $4*4$ spectral patch corresponding to the original spectral space. These spectral patches can be regarded as instances of acoustic events, and classification frameworks based on bags of multiple instances can be applied. In most of the frameworks, average feature representations obtained from the last top layer are used for the classification \cite{SuICASSP17, LiuACM16, Mardon1998, Wang2018,Ramon2000,Kumar2017,Tseng2018}. Local small-scale representations in other lower layers, which encode discriminative information, may be smoothed out. We believe that such discriminative information should be further utilized, i.e., propagating multi-scale representations from the bottom to top layers when performing classification.
\subsection{Proposed  CSA model}
\label{sec-II-II}
\begin{figure}[tb]
\centering
\includegraphics[width=7cm, height=3.5cm]{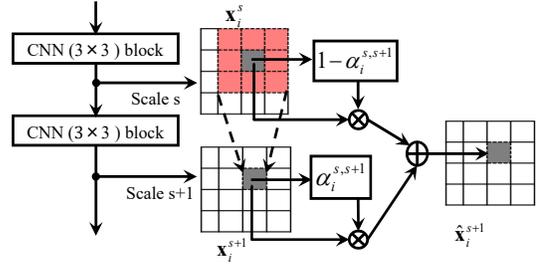}
\caption{Attention weighting in consecutive scales in a DCNN (with a CNN kernel as $3 \times 3$).}
\label{fig2}
\end{figure}

In multi-layer CNNs, multi-scale features are extracted and propagated in consecutive layers, and their importance can be measured by a scale attention model. A direct realization of this scale attention is to adaptively weight multi-scale features in consecutive layers, as illustrated in Fig. \ref{fig2}. In this figure, the convolutional block is formed by a ``CNN ($3 \times 3$) block,"  each comprising a linear convolutional operator and a nonlinear transform. The output of the layer with scale $s+1$ has a larger receptive field than that of the layer with scale $s$ (output of the preceding layer). Each ``pixel" in the layer with scale $s+1$ can be regarded as a ``smoothed" process from a ``patch" ($3 \times 3$) in the preceding layer with scale $s$. To keep the discriminative structure in the consecutive layers, the output should be a weighted combination based on their importance (discrimination ability in classification tasks) from the two feature spaces. For illustration and explanation, we show the attentive weighting from the two consecutive scales only in one filter channel, as presented in Fig. \ref{fig2}. In Fig. \ref{fig2}, the transform between the two consecutive scales $s$ and $s+1$ is defined as:
\begin{equation}
 {\bf x}_i^{s + 1}  = h^{s,s + 1} \left( {{\bf x}_i^s } \right),
 \label{eq2}
 \end{equation}
 where $h^{s,s + 1} \left(  \cdot  \right)$ is the feature transform between scale $s$ and scale $s+1$.  It is the transform function of the convolution block as showed in Fig. \ref{fig2}. In each ``pixel" position, the attended output is represented as:
\begin{equation}
{\bf \hat x}_i^{s + 1}  = \left( {1 - \alpha _i^{s,s + 1} } \right){\bf x}_i^s  + \alpha _i^{s,s + 1} {\bf x}_i^{s + 1}
\label{eq3}
\end{equation}
where ${\alpha _i^{s,s + 1} }$ is the attention weight on the $i$-th position (in ``pixel" level) between scale $s$ and $s+1$, and it is constrained as $0 \le \alpha _i^{s,s + 1}  \le 1$. Eq. \ref{eq3} shows that in two consecutive scales, the output in each position is a weighted summation based on their importance in the two scales. Eq. \ref{eq3} is further cast to:
\begin{equation}
{\bf \hat x}_i^{s + 1}  = {\bf x}_i^s  + \alpha _i^{s,s + 1} \left( {{\bf x}_i^{s + 1}  - {\bf x}_i^s } \right)
\label{eq4}
\end{equation}
From Eq. \ref{eq4}, we can see that the term $\left( {{\bf x}_i^{s + 1}  - {\bf x}_i^s } \right)$ is exactly the same as the residual branch used in the residual network (ResNet) \cite{HeCVPR2016, HeECCV2016}. In the following, we illustrate our idea with a connection to the ResNet, which is one of the most widely used state-of-the art model frameworks in deep learning.

\subsection{ResCNN}
\label{sec-II-III}
\begin{figure}[tb]
\centering
\includegraphics[width=7cm, height=3.5cm]{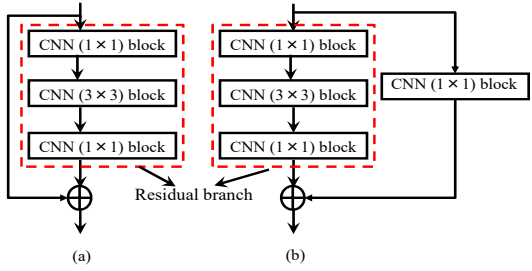}
\caption{ResNet (CNN): (a) Identity residual block, and (b) Convolutional residual block.}
\label{fig_resnet}
\end{figure}

The original purpose of using the ResNet architecture \cite{HeCVPR2016, HeECCV2016}  is to deal with the vanishing gradient problem when the deep network is optimized based on gradient back-propagation (BP) algorithms \cite{BP1986}. By using skip connections, the gradient that is used for model parameter updating can be efficiently back-propagated to the lower layers in a deep network. Therefore, model parameters for the top and bottom layers can be well trained. Besides the original motivation for the proposal of the ResNet, several studies have revealed that the benefit of the ResNet is more than making the training of a deep network efficient; it can also lead to a good model generalization due to the ensemble and stochastic depth properties of the model structure \cite{AndreasNIPS2016, HuangECCV2016}. When convolutional blocks are used in the ResNet, small-scale features can be carried from the bottom to top layers through skip connections. There are two types of ResNet blocks with convolution operators (ResCNN) (Fig. \ref{fig_resnet}), one is with an identity connection (panel (a)), and the other is with a convolutional connection to make the input and output compatible in feature dimensions (panel (b)). In this figure, the convolutional block includes a batch normalization (BN) \cite{BNNorm} and a nonlinear activation function, which are omitted in this paper for convenience of explanation. The transform in this ResCNN block is:
\begin{equation}
{\bf x}^{s + 1}  = {\bf x}^s  + f_{\rm res}^{s,s + 1} \left( {{\bf x}^s } \right),
\label{eq5}
\end{equation}
where ${\bf x}^s$ is the input of the ResCNN block, ${\bf x}^{s + 1}$ is the output, and $f_{\rm res}^{s,s + 1} \left(  \cdot  \right)$ is a transform function of the residual branch.

\subsection{Relationship between CSA and ResCNN}
\label{sec-II-IV}
\begin{figure}[tb]
\centering
\includegraphics[width=8cm, height=4cm]{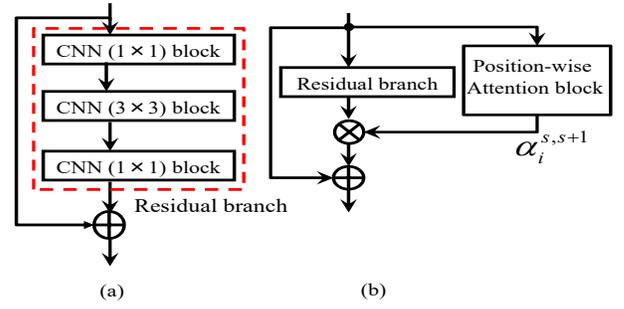}
\caption{A comparison of ResCNN  and the propsoed CSA model : (a) ResCNN (a special case with $\alpha _i^{s,s + 1}=1$), (b) The CSA model with a position-wise scale attention.}
\label{fig_resatt}
\end{figure}

From Eq. \ref{eq5}, we can see that the ResCNN tries to learn the residual transform function as:
\begin{equation}
f_{\rm res}^{s,s + 1} \left( {{\bf x}^s } \right) = {\bf x}^{s + 1}  - {\bf x}^s
\label{eq6}
\end{equation}
By substituting Eq. \ref{eq6} in Eq. \ref{eq4}, we obtain the scale attention model as:
\begin{equation}
{\bf \hat x}_i^{s + 1}  = {\bf x}_i^s  + \alpha _i^{s,s + 1} f_{\rm res}^{s,s + 1} \left( {{\bf x}_i^s } \right)
\label{eq7}
\end{equation}
Here, the proposed CSA model can be regarded as an extension of the conventional ResCNN. The conventional ResCNN is actually an implicit multi-scale feature processing that propagates small-scale features through the skip connections, i.e., with $\alpha _i^{s,s + 1}=1$. In the transform function of the residual branch, $f_{\rm res}^{s,s + 1} \left(  \cdot  \right)$ has a $3*3$ convolution operation. By directly adding features in scale $s$ to the output, the residual branch tries to learn more on large-scale features in scale $s+1$. Based on this analysis, we assume that the benefit of using the ResCNN is the efficient propagation of multi-scale features in feature representation, which helps to integrate local and global feature dependencies for classification. Our CSA model can be implemented with the structure of the ResCNN as a backbone by adding attention blocks on the residual branches (as shown in panel (b) of Fig. \ref{fig_resatt}).

\subsection{Design of the CSA function as a feed-forward attention network}
\label{sec-II-V}
\begin{figure}[tb]
\centering
\includegraphics[width=8cm, height=4cm]{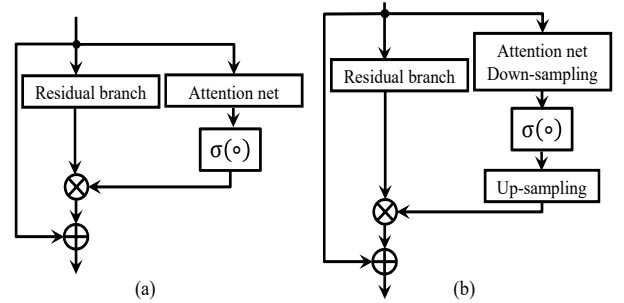}
\caption{The CSA blocks without down-up sampling (a), and with down-up sampling (b).}
\label{figall_sub2}
\end{figure}

There are many strategies for the estimation of the attention weight $\alpha _i^{s,s + 1}$ as used in Eq. \ref{eq7}, for example, feed-forward attention, feed-back attention, and local and global attentions \cite{Hahdanau14,Luong15, Colin2015}. In our study, as a general form, a feed-forward attention network is designed. The position-wise attention weight for features between scales $s$ and $s+1$ is defined as:
\begin{equation}
\alpha _i^{s,s + 1}  = \sigma \left( {f_{{\rm att\_net}}^{s,s + 1} \left( {{\bf x}_i^s ,context\left( {{\bf x}_i^s } \right),{\bf W}_{{\rm att\_net}}^{s,s + 1} } \right)} \right)
\label{eq8}
\end{equation}
 In this equation, $f_{{\rm att\_net}}^{s,s + 1} \left(  \cdot  \right)$ is a network function that takes ${{\bf x}_i^s }$ and its context ${context\left( {{\bf x}_i^s } \right)}$ as inputs, and ${{\bf W}_{{\rm att\_net}}^{s,s + 1} } $ is the attention network parameters. In our study,  $f_{{\rm att\_net}}^{s,s + 1} \left(  \cdot  \right)$ is implemented as a CNN, the context information of ${{\bf x}_i^s }$ can be easily integrated via convolutional operators. $\sigma \left(  \circ  \right)$ is the logistic sigmoid function to constrain the attention value in the range $\left[ {0,1} \right]$. Considering the different scale resolution in attention weighting, two types of implementations are proposed, one is that the attention net keeps the same resolution as that of the residual branch (in panel (a) of Fig. \ref{figall_sub2}), the other is that the attention network is with down- and up-sampling (in panel (b) of Fig. \ref{figall_sub2}). In our experiments, we will investigate these two types of implementation. From Eq. \ref{eq8}, we can see that the CSA is actually an adaptive weighting process depending on current input and its context. This is another important feature that differs from the ResCNN, which uses a fixed-scale feature summation in consecutive scales. Meanwhile, this scale attention is actually a general and unified form of attention modeling because temporal and spatial information can be integrated in the calculation as context information.

\section{Implementation of the CSA model for AEC}
\label{sec-III}
\begin{figure}[tb]
\centering
\includegraphics[width=8cm, height=3.5cm]{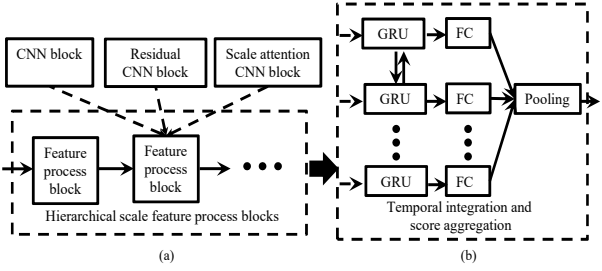}
\caption{The proposed framework of the CSA model for AEC: (a) feature process module, and (b) temporal integration and score aggregation module.}
\label{figall}
\end{figure}

With the prepared definitions and explanations in Subsections \ref{sec-II-II}, \ref{sec-II-III}, \ref{sec-II-IV}, and \ref{sec-II-V}, we propose a computational framework, as shown in Fig. \ref{figall}.
This framework follows the state-of-the-art pipeline for the AEC task, where invariant feature extraction and temporal context integration are modeled with the CNN and RNN \cite{Emre17, ChoiICASSP17}, respectively. 
\subsection{CSA with hierarchical scale and temporal integration}
There are two modules involved in the framework: one is concerned with hierarchical-scale feature process, and the other is related to the temporal integration and score aggregation process. The feature process blocks in the first module can be realized by CNN blocks, ResCNN blocks, or our proposed CSA blocks. The explored features are aggregated to give classification scores in the final stage with a pooling operation. The design of the two modules are introduced in detail in the following.
\subsubsection{Implementation of the CSA block}
\label{sec-II-subsub}
\begin{figure}[tb]
\centering
\includegraphics[width=8cm, height=2.5cm]{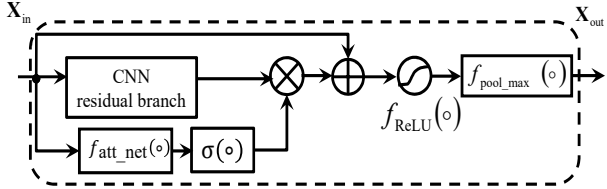}
\caption{Implementation of the CSA block.}
\label{figall_sub1}
\end{figure}

Based on our analysis, the scale attention can be formulated as a position-wise attention on a residual branch where the ResCNN is used as a backbone. The implementation of the scale attention block is shown in Fig. \ref{figall_sub1}. By stacking several-scale attention blocks in a feature process module (Fig. \ref{figall}), we can obtain the refined cross-scale feature representation of spectral patches, which will be used in the next stage for AEC.

\subsubsection{Temporal integration and score aggregation}
\begin{figure}[tb]
\centering
\includegraphics[width=6cm, height=5cm]{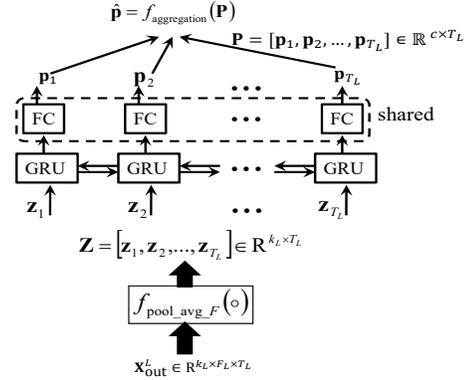}
\caption{Temporal integration and score aggregation for class probability estimation.} \label{figGR}
\end{figure}

After obtaining importance weighted features from the CSA model, the temporal integration and score aggregation module is designed, as shown in Fig. \ref{figGR}.
In this figure, the input ${\bf Z} = \left[ {{\bf z}_1 ,{\bf z}_2 ,...,{\bf z}_{T_L } } \right] \in {\mathbb R}^{k_L  \times T_L }$ is the representation processed by the feature process module (${\bf x}_{{\rm out}}^L$ and the output of the last feature extraction module from panel (a) of Fig. \ref{figall}) with an average pooling along the frequency axis (as operator $f_{{\rm pool\_avg\_}F} \left(  \circ  \right)$ in the figure). A bidirectional recurrent network (BGRU units are used) is applied to further process the sequence. A fully connected (FC) layer with a softmax activation is stacked in each output of the BGRU process. The output is a probability vector. The process in this block is formulated as follows:
\begin{equation}
\begin{array}{l}
 {\bf \mathord{\buildrel{\lower3pt\hbox{$\scriptscriptstyle\leftrightarrow$}}
\over Z} } = {\rm BGRU}\left( {\bf Z} \right) = \left[ {{\bf \mathord{\buildrel{\lower3pt\hbox{$\scriptscriptstyle\leftrightarrow$}}
\over z} }_1 ,{\bf \mathord{\buildrel{\lower3pt\hbox{$\scriptscriptstyle\leftrightarrow$}}
\over z} }_2 ,...,{\bf \mathord{\buildrel{\lower3pt\hbox{$\scriptscriptstyle\leftrightarrow$}}
\over z} }_{T_L} } \right] \in {\mathbb R}^{k_L \times T_L}  \\
 {\bf o}_i  = {\bf W}_{\rm cls} {\bf \mathord{\buildrel{\lower3pt\hbox{$\scriptscriptstyle\leftrightarrow$}}
\over z} }_i  + {\bf b}_{\rm cls}  \in {\mathbb R}^c  \\
 p_{i,j}  = \frac{{\exp \left( {o_{i,j} } \right)}}{{\sum\limits_j {\exp \left( {o_{i,j} } \right)} }} \\
 {\bf p}_i  = \left[ {p_{i,1} ,p_{i,2} ,...,p_{i,c} } \right] \in {\mathbb R}^c  \\
 {\bf P} = \left[ {{\bf p}_1 ,{\bf p}_2 ,...,{\bf p}_{T_L} } \right] \in {\mathbb R}^{c \times T_L}  \\
 {\bf \hat p} = f_{{\rm aggregation}} \left( {\bf P} \right) \in {\mathbb R}^c,  \\
 \end{array}
\label{eq14}
\end{equation}
where ${\bf W}_{\rm cls}$ and ${\bf b}_{\rm cls}$ are the weight matrix and bias of the FC layer, respectively, ${\bf \mathord{\buildrel{\lower3pt\hbox{$\scriptscriptstyle\leftrightarrow$}}
\over z} }_i$ is the output of the $i$-th step from the BGRU layer, and $c$ is the number of classes. The final bag-level probability vector is obtained with an aggregation function $f_{{\rm aggregation}} \left(  \circ  \right)$. In this study, an average pooling is used as the aggregation function.

We can formulate the estimated class probability vector by a function of the input acoustic spectrum and neural network parameters as ${\bf \hat p}_m = {\bf \hat p}\left( {{\bf X}_m^0 ;\Theta } \right)$ (estimated from Eq. \ref{eq14}), where ${\bf X}_m^0$ is an input acoustic sample (raw input spectrum at scale $0$), $m$ is the sample index as $m=1,2,...,N$ ($N$ is the total number of samples), and $\Theta$ is the network parameter set. The learning is based on minimizing a loss function defined as the cross-entropy (CE) of the predicted and true targets as:
\begin{equation}
L_{{\rm CE}}  =  - \frac{1}{N}\sum\limits_{m = 1}^N {{\bf y'}_m \log {\bf \hat p}_m \left( {{\bf X}_m^0 ;\Theta } \right)},
\label{eq15}
\end{equation}
where ${\bf y'}_m$ is the transpose of the true target label vector ${\bf y}_m$. In most studies, a parameter $L_2$ regularization ($\left\| .\right\|_2^2$) defined as in Eq. \ref{eq16} is added in the objective function with a trade-off weighting coefficient.
\begin{equation}
L_{{\rm l2\_reg}} \left( \Theta  \right) = \left\| \Theta  \right\|_2^2.
\label{eq16}
\end{equation}

\subsection{Further discussion on the implementation of the attention network}
\label{sec-III-I}
In the previous subsection and Subsection \ref{sec-II-V}, we have explained the scale attention as a position-wise attention on the residual branch. In this subsection, we further examine its design details with regard to the temporal and spatial attentions in feature weighting. Intuitively, as a sequential perception problem, the perception importance can be allocated to some temporal frames or segments, i.e., the attention variable is a 1D time-dependent variable. In this study, however, we defined acoustic event patterns as representations of temporal-frequency patches. Therefore, the output of the attention function can be a 2D spatial-dependent variable. Furthermore, in the 2D CNN framework, the output of each layer is a 3D tensor, e.g., the $l$-th layer output ${\bf X}^l  \in {\mathbb R}^{k_l  \times F_l  \times T_l }$ has a feature channel index besides the two spatial dimensions. In this situation, the CNN filters are regarded as 2D instance detectors, and each feature channel can be associated with a 2D attention weight matrix. In this sense, the output of the attention model should be a 3D channel spatial-dependent variable. Theoretically, 1D (time) attention model can be regarded as a special case of a 2D (spatial) attention model with an average along the frequency dimension, whereas a 2D attention model is a special case of a 3D (channel-spatial) attention model with an average along the channel dimension. Although models with a high representation capacity should outperform those with a low representation capacity, in real applications, we need to consider the effect of model complexity. In the following, we show how to estimate the attention functions for each of these situations.

\begin{figure}[tb]
\centering
\includegraphics[width=4cm, height=5cm]{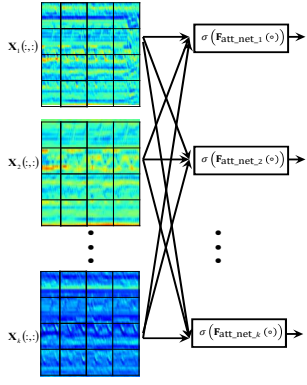}
\caption{Cross-channel spatial attention model (CC\_SAM\_3D).}
\label{fig3DAtt}
\end{figure}

We suppose that an acoustic event is finally represented as a 3D tensor ${\bf X}^{3D} \in {\mathbb R}^{k \times F \times T}$, where $k,F, T$ are the channel index and two spatial dimensions of the height and width, respectively. In each channel (corresponding to index $k$), there is an attention matrix to assign weights to each ``pixel." This 3D attention model is called the cross-channel spatial attention model (CC\_SAM\_3D), as shown in Fig. \ref{fig3DAtt},
The attention map function for the $i$-th channel is designed as:
\begin{equation}
{\bf A}_i  = \sigma \left( {{\bf F}_{{\rm att\_net\_}i} \left(  \circ  \right)} \right),
\end{equation}
where ${\bf A}_i  \in {\mathbb R}^{F \times T}$ is a 2D attention matrix (the consecutive scale indexes are omitted here for easy explanation) and ${{\bf F}_{{\rm att\_net\_}i} \left(  \circ  \right)}$ is an attention network transform (before the sigmoid activation) for the $i$-th channel. 

\subsection{Orthogonality constraint on the attention matrix}
\label{orthogonal}
Apparently, in the 3D attention model, different filter channels may pay attention to the same local features of spectral patches. To constrain the output of each filter channel and focus on different local features, an attention map orthogonality constraint is added in the optimization. Correspondingly, the attention orthogonality loss function is defined as:
\begin{equation}
L_{{\rm ortho\_reg}}  = \left\| {{\bf M}'{\bf M} \otimes \left( {{\bf 1} - {\bf I}} \right)} \right\|_F^2,
\label{eq26}
\end{equation}
where $\otimes$ is an element-wise multiply operator, $\left\|  \circ  \right\|_F^2$ is the Frobenius norm, ${\bf M}'$ is the transpose of $\bf M$, and each column of $\bf M$ is a flattened attention matrix of a filter channel ${\bf A}_i$, ${\bf 1}$ is an all-ones square matrix, and ${\bf I}$ is an identity matrix. The loss defined in Eq. \ref{eq26} is an orthogonal constraint of the attention, i.e., each filter channel tries to catch uncorrelated attention components. The final optimization is based on minimizing the following objective function:
\begin{equation}
L_{{\rm net}}  = L_{{\rm CE}}  + \frac{{\lambda _1 }}{2}L_{{\rm l2\_reg}} \left( \Theta  \right) + \frac{{\lambda _2 }}{2}L_{{\rm ortho\_reg}},
\label{eq27}
\end{equation}
where $L_{{\rm l2\_reg}} \left( \Theta  \right)$ is defined in Eq. \ref{eq16} and $\lambda _1$ and $\lambda _2$ are two regularization parameters.

\subsection{Specialized CSA models}
Based on the 3D attention model, special cases of the CSA models can be obtained as follows.
\subsubsection{Channel-wise spatial attention model}
\label{parsemi3D}
\begin{figure}[tb]
\centering
\includegraphics[width=4cm, height=4.5cm]{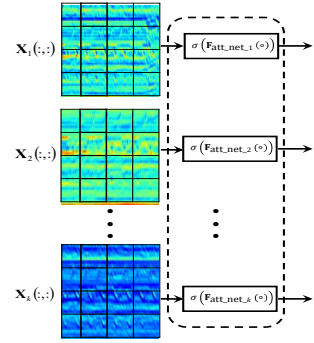}
\caption{Channel-wise spatial attention model (CW\_SAM\_2.5D).} \label{figSemi3DAtt}
\end{figure}

Different from the cross-channel spatial attention model, a channel-wise 2D spatial attention matrix (pseudo 3D, which is denoted as CW\_SAM\_2.5D in this paper) is estimated independently for each feature channel. There are two types of designs considering whether their model parameters are shared or not. The design is presented in Fig. \ref{figSemi3DAtt}.
In this figure, there is no cross connections between different feature channels in the attention network design, and thus this system can be considered as an intra-channel attention model. With different inputs from each channel, different attention maps for each channel can be obtained.
\subsubsection{Spatial attention model}
\label{subsub2D}
\begin{figure}[tb]
\centering
\includegraphics[width=5cm, height=5.5cm]{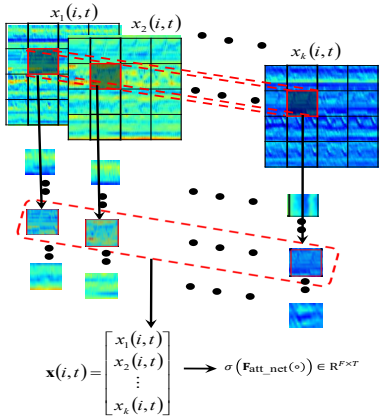}
\caption{All channels share the same spatial attention map (SAM\_2D).}
\label{fig2DAttAll2One}
\end{figure}

As acoustic event patterns are encoded in time–frequency spectral patches, a 2D spatial attention model (SAM\_2D) can be applied naturally. In addition, in contrast to the pseudo 3D attention model, all channels share the same one attention matrix. The design is shown in Fig. \ref{fig2DAttAll2One}.
In this design, each spectral patch extracted from the spatial location $\left( {i,t} \right)$ (as a ``pixel" in the output space of the CNN layer) is represented as ${\bf x}\left( {i,t} \right) \in {\mathbb R}^k$ ($F \times T$ for an input acoustic signal). It is represented as a collection of representations across all filter channels. If spatial contexts are taken into consideration, then the concatenation of features from spatial context is applied.
\subsubsection{Temporal attention model}
\label{subsub1D}
As a temporal sequence, spectral patches share the same importance if they are extracted from the same time stamp in the original spectral space. This is the most intuitive application of the attention model in AEC. In the design, based on the CC\_SAM\_3D, CW\_SAM\_2.5D, and SAM\_2D attention models, in each temporal location, we obtain a temporal attention weight by an average along the frequency dimension. Correspondingly, there are many types of designs for the temporal attention model, for example, cross-channel temporal attention model (CC\_TAM\_1D), channel-wise temporal attention model (CW\_TAM\_1D), and one temporal attention map model (TAM\_1D) shared by all channels. In this study, only the results based of TAM\_1D are investigated.

\subsubsection{Architecture of the attention model}
The attention function can be implemented as an FC  neural network or a locally connected (LC) CNN. Due to the flexibility of the CNN, in this study, the LC-based CNN was implemented for attention network models. In the attention model with independency assumption for spectral patches, $1 \times 1$ convolution kernel is used. Meanwhile, with the dependency assumption, the kernel size is set with a given spatial neighborhood size ($3 \times 3$ neighborhood size is used in this study), i.e., the attention for one patch is estimated based on the patches surrounding the focused patch in spatial locations.

From the analysis above, we can see that in most studies using the attention models for AEC, for example, the temporal attention model or multiple-instance attention model \cite{KongICASSP17, XuIS17, LuIS2018,Ilse2018}, their algorithms can be regarded as a special case of our CSA model. To the best of our knowledge, we are the first to reveal the relationship between the CSA model and ResCNN and generalize the CSA model for AEC.

\section{Experiments}
\label{sec-IV}
Two data corpora are adopted to test the proposed framework: one is the UrbanSound8K corpus for acoustic event and scene classification \cite{Salamon2017,Salamon2014}  and the other is the data set of DCASE 2017 task 4 (DCASE2017\_T4) for large-scale weakly supervised acoustic event detection in smart car environments \cite{DCASE2017}. UrbanSound8K consists of 8,732 sound clips (less than 4 seconds) of 10 classes labeled on the clip level as air conditioner (AC), car horn (horn), playing children (child), barking dogs (dog), drilling (drill), engine idling (engine), gunshots (gun), jackhammer (hammer), siren, and street music (music). All sounds were organized into 10 folds. In our study, each of the 10 folds is alternatively selected as the test set, and the remaining nine folds are utilized as training and validation sets (validation set is selected from one of the nine folds with the remaining taken as the training set alternatively (\cite{Salamon2017} )). The final evaluation is based on an average performance on the 10 test fold sets. As the sounds were recorded and collected from crowd sources, and labels were given only on the clip level without accurate start and ending time stamps, the classification of these sounds is difficult and realistic. The DCASE2017\_T4 data set is a subset of Google Audioset \cite{Audioset}. It consists of 17 types of sound events that occur in car or vehicle environments. Each sample is collected as a clip with a duration around or less than 10 seconds. In total, there are 51,172 clips for training, 488 clips for testing (used as the validation set for selecting good models for the final evaluation), and 1,103 clips for the final evaluation \cite{DCASE2017}. As a weak-label classification task, only tag label information on the clip level is applied. In our study, we first utilize the UrbanSound8K corpus for a detailed analysis and model comparisons as many studies have carried out research work based on deep architectures to test their algorithms on this particular data set \cite{Salamon2017,Piczak2015,Salamon2014}, and then we carry out experiments on the DCASE2017\_T4 corpus to further verify the proposed framework.
\subsection{Implementation details}
The raw input feature to the model is log-compressed Mel spectrogram (MSP). In the MSP extraction, all sounds were down-sampled with a 16 kHz sampling rate. A 512-point windowed fast Fourier transform  with a 256-point shift was used for the frame-based power spectrum extraction, and 60 Mel filter bands were used for the MSP representation. As our models take spectral patches in convolution for feature processing, a 2D MSP feature matrix for each event clip was used as the input to the models.

Many models based on deep learning algorithms have been proposed to improve the performance of the AEC task \cite{Salamon2017,Piczak2015,DCASE16, Aytar2016, Lishao2018}. Among these models, the DCNN-based models perform consistently well due to their strong power in temporal-frequency invariant feature extraction. With the incorporation of temporal context modeling by an RNN layer (with either LSTM or GRU units), the CRNN always gives the state-of-the-art performance in most studies related to audio classification and detection \cite{Emre17, ChoiICASSP17}. In our study, as a baseline model, we implemented the deep CRNN (DCRNN) model for comparison. In addition, as an implicit multi-scale feature integration model, the ResCNN-based model is also implemented. With the ResCNN as our backbone model, our scale attention models were built. Besides the various selection of model architectures, many methods or techniques can be applied to improve the AEC performance, for example, data augmentation, optimization algorithms, transfer learning, cross-modal learning, or rovering of many sub-systems. However, comparing the results with different methods sometimes is not fair or cannot help in understanding the internal problems because we need a lot of tricks to tune a model from many aspects. In our study, with a very typical architecture and learning algorithm, we only focus on whether integrating the CSA model can improve the discriminative feature extraction for improving the performance or not. Therefore, all processing procedures follow the same pipeline for a fair comparison, which is shown in Fig. \ref{figall}. For testing different models, only the feature process blocks are replaced as plug and play. In the CNN blocks for local feature extraction, each block includes one linear CNN layer (with 256 of 3*3 filter kernels), followed by a BN operation (for improving the convergence speed and model generalization), a nonlinear activation (rectified linear unit [ReLU] was used in this study), and a max-pooling operation (with 2*2 strides) (as defined in Eq. \ref{eq1}). After feature extraction in the DCNN, a bidirectional RNN layer is applied on the output of the last CNN block for temporal context modelling, followed by an FC layer and softmax layer for classification. In the RNN layer, 128 GRU nodes (to obtain 256-dimension event feature vectors) were used (as shown in Fig. \ref{figGR}). In the implementation of the ResCNN, the identity residual block shown in panel (a) of Fig. \ref{fig_resnet} is used. In the residual block, except the last CNN block, each CNN block has a linear CNN layer, BN operation, and ReLU activation. Only the last CNN block has a linear CNN layer, and the BN and nonlinear activation are applied after the identity addition. Moreover, in each residual block, the output dimension is 256, and the bottleneck compression ratio is 4. The CSA model applied the ResCNN as a backbone with a modification of each block, following the design in Fig. \ref{figall_sub1}. In the model training, the regularization coefficient $\lambda _1$ (in Eq. \ref{eq27}) in the objective function was fixed to $0.0001$, and $\lambda _2$ was experimentally decided. A mini-batch size 32 was set, and a stochastic optimization Adam algorithm with a learning rate $0.001$ was applied \cite{Adam} in the model optimization learning.
\subsection{Experiments on the UrburnSound8K corpus}
Before carrying out the experiments the for final evaluations, we first performed experiments on the UrbanSound8K corpus to confirm several factors that affect the performance.
\subsubsection{Baseline models}
To select good baseline models, we tried several configurations of ``DCRNN" with different layers of CNN blocks. The results are shown in Table \ref{tab1}.
\begin{table}[tb]
\centering
\caption{Performance of baseline models (classification accuracy (standard deviation)) (\%)}
\begin{tabular}{|c||c|c|}
\hline
 Methods&Valid&Test\\
\hline
\hline
DCRNN (1 CNN block)&74.8 (4.60) & 72.1 (4.95)\\
\hline
DCRNN (2 CNN blocks)&78.2 (5.10) &74.8 (5.34)\\
\hline
DCRNN (3 CNN blocks)&79.4 (5.52) &\textbf{76.2} (6.01)\\
\hline
DCRNN (4 CNN blocks) &79.1 (5.54) &75.1 (5.51)\\
\hline
PiczakCNN (\cite{Piczak2015}) & - &73.7 (-) \\
\hline
SB-CNN (\cite{Salamon2017}) & - &73.0 (-) \\
\hline
\end{tabular}
\label{tab1}
\end{table}
The table present the classification accuracies and their standard deviations summarized on the 10 test folds. Compared with the results of the test set, the model with one convolutional block does not have sufficient model capacity, which results in underfitting. With the increase of CNN processing blocks, the performance is consistently improved. However, adding more convolutional blocks beyond three has no further benefit on the AEC task. Compared with other studies on this data corpus, for example, the DCNN models in \cite{Piczak2015, Salamon2017}, the ``DCRNN" model with three feature processing blocks is a more reasonable baseline model.

\begin{figure}[tb]
\centering
\includegraphics[width=9cm, height=4cm]{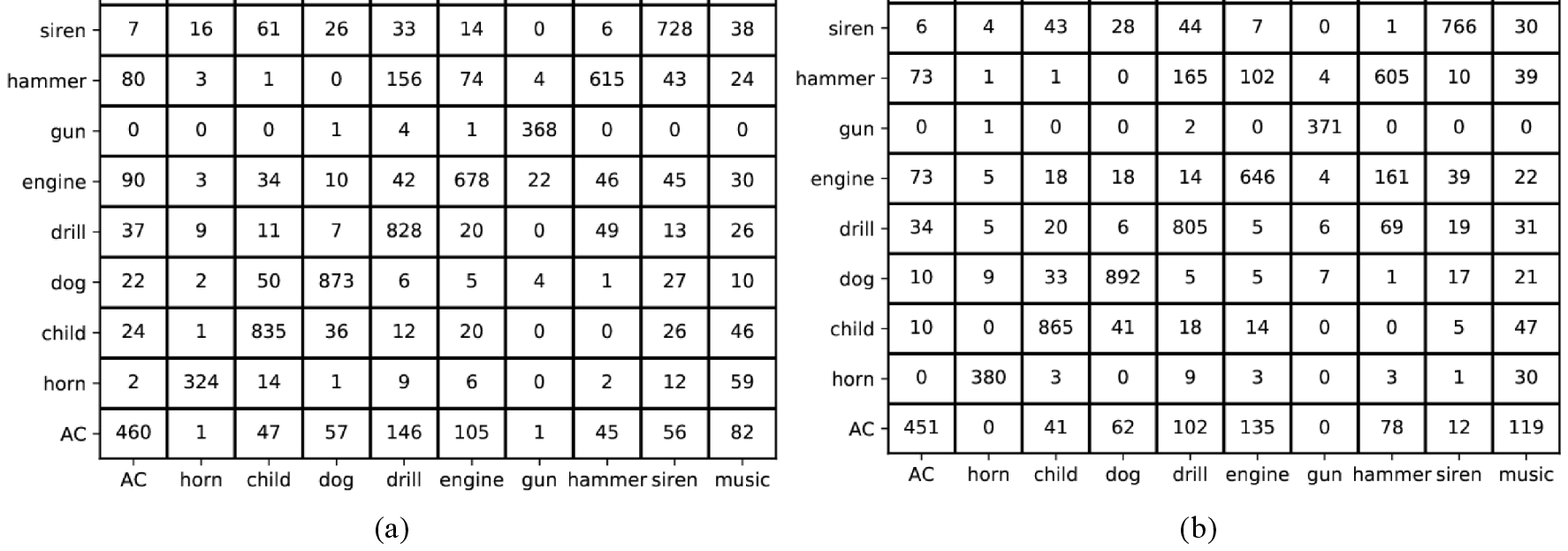}
\caption{Confusion matrix on UrburnSound8K corpus for the DCRNN baseline models with different layers of CNN blocks, (a) two CNN blocks, (b) three CNN blocks.}
\label{Confusion matrix}
\end{figure}

In the baseline models with different layers of CNN blocks, their features for acoustic event discrimination can be regarded from different scales. For a detailed analysis, we show the confusion matrix of acoustic events based on the models with two and three CNN blocks in Fig. \ref{Confusion matrix}.
This confusion matrix shows that although the total performance with three CNN blocks (panel (b)) is improved compared with that with two CNN blocks (panel (a)), the improvement has performance degradation cost on some specific event categories. For example, as shown in Fig. \ref{Confusion matrix}, the accuracies of recognizing events ``AC," ``hammer," ``engine," and ``drill" are decreased (from panel (b) to panel (a)). Moreover, although the confusion from ``horn" to ``music" is reduced when features are represented in a large scale, the confusion from ``AC" to ``music" is increased. For a further investigation, in Table \ref{tab2}, we show the performance changes when the number of layers of CNN blocks is changed.
\begin{table}[tb]
\centering
\caption{Effect of changing number of CNN layers in the DCRNN model (\%)}
\begin{tabular}{|c||c|c|c|c|}
\hline
 Layers&1 $\Rightarrow$ 2 &2 $\Rightarrow$ 3 &3 $\Rightarrow$ 4\\
\hline
\hline
Corrected&10.9 & 7.5&4.4\\
\hline
Degraded&7.4 &6.1&5.5\\
\hline
Kept&63.9 &68.7&70.7\\
\hline
\end{tabular}
\label{tab2}
\end{table}
In this table, the columns with ``Corrected," ``Degraded," and ``Kept" are defined as:
\begin{equation}
\begin{array}{l}
 {\rm Corrected}\left( {{\rm M1} \Rightarrow {\rm M2}} \right){\rm   = }\frac{{\# \left\{ {\left( {{\rm Wrong|M1}} \right) \cap \left( {{\rm Correct|M2}} \right)} \right\}}}{{{\rm \# Total }}} \\
 {\rm Degraded}\left( {{\rm M1} \Rightarrow {\rm M2}} \right){\rm   = }\frac{{\# \left\{ {\left( {{\rm Correct|M1}} \right) \cap \left( {{\rm Wrong|M2}} \right)} \right\}}}{{{\rm \# Total }}}  \\
 {\rm Kept}\left( {{\rm M1} \Rightarrow {\rm M2}} \right){\rm   = }\frac{{\# \left\{ {\left( {{\rm Correct|M1}} \right) \cap \left( {{\rm Correct|M2}} \right)} \right\}}}{{{\rm \# Total }}} \\
 \end{array}
\end{equation}
where ${\# \left\{ {\left( {{\rm Wrong|M1}} \right) \cap \left( {{\rm Correct|M2}} \right)} \right\}}$ is the number of samples of correctly predicted by model $\rm M2$ and wrongly predicted by model $\rm M1$. Other definitions follow similar meanings. Accordingly, we can know when the number of CNN layers is changed (e.g., from model M1 to model M2) and what percentage of clips are corrected, degraded, and kept during recognition. As shown in Table \ref{tab2}, in the baseline model, with the increasing number of CNN blocks, there is a tendency that the number of corrected clips is larger than that of degraded clips in recognition. However, the tendency is reverse when the model changes from three CNN blocks to four blocks. In summary, most convention models stack several CNN layers try to find an optimal scale of feature representation, based on which the number of corrected patterns is larger than that of degraded patterns.
\subsubsection{ResCNN and CSA models}
We test the performance of the ResCNN model and the proposed CSA models in which the ResCNN is used as a backbone. In all these models, three feature processing blocks are adopted based on the initial analysis of baseline models. According to the same processing pipeline in Fig. \ref{figall}, we replace the CNN blocks with ResCNN blocks and the CSA blocks in the feature processing module. In a common usage of ResNet \cite{HeCVPR2016, HeECCV2016}, the number of neural nodes is increased from the bottom to top layers. For example, the output nodes for residual blocks can be arranged as 64-256-512 in a ResNet with three residual blocks. However, as our empirical experiments show, a plain usage of residual blocks with 256-256-256 obtained slightly better results. Moreover, for a fair and consistent comparison with using CNN blocks for feature extraction, in our implementations, the identity convolutional blocks were used. Our scale attention model also follows the same configuration based on the ResCNN and attention map orthogonal regularization $\lambda _2 =0.1$ (based on our empirical experiments). In addition, in the implementation of the CSA model, as shown in Fig. \ref{figall_sub2}, attention networks with and without down- and up-samplings are designed. Three down-sampling methods are examined, i.e., CNN stride convolution, average pooling, and max pooling. All of these down-samplings have 2*2 strides. The bilinear interpolation was used in the up-samplings. The performance of these models are shown in Table \ref{tab3}.
\begin{table}[tb]
\centering
\caption{Performance of ResCNN and the cross-channel-based CSA models (classification accuracy (standard deviation)) (\%)}
\begin{tabular}{|c||c|c|}
\hline
 Methods&Valid&Test\\
\hline
\hline
ResCNN &80.5 (5.56) &77.5 (5.69)\\
\hline
\hline
CC\_SAM\_3D (No sampling)&81.9 (5.75)&79.0 (6.01)\\
CC\_SAM\_3D (CNN down-up)&81.7 (4.76)&79.2 (4.34)\\
CC\_SAM\_3D (AP down-up)&81.3 (5.34)&78.8 (5.63)\\
CC\_SAM\_3D (MP down-up)&81.9 (5.19)&\textbf{79.4} (6.29)\\
\hline
\end{tabular}
\label{tab3}
\end{table}

\begin{table}[tb]
\centering
\caption{Performance of channel-wise, spatial, and temporal-based CSA models (classification accuracy (standard deviation)) (\%)}
\begin{tabular}{|c||c|c|}
\hline
 Methods&Valid&Test\\
\hline
\hline
CW\_SAM\_2.5D&81.5 (5.24)&\textbf{79.2} (5.60)\\
CW\_SAM\_2.5D (shared)&81.8 (5.49)&78.8 (5.77)\\
SAM\_2D&80.9 (5.19) &78.6 (5.50)\\
TAM\_1D &80.7 (5.40)&78.2 (5.59)\\
\hline
\end{tabular}
\label{tab3_2}
\end{table}
The comparison of the results in Tables \ref{tab1} and \ref{tab3} show that the ResCNN significantly improved the performance compared with the model using CNN blocks without scale feature skip connections. The CSA models can further improve the performance of the ResCNN model. Moreover, the CSA models with the down- and up-samplings slightly improved the performance (except with the average pooling-based down-sampling method). Based on these results, in future experiments and analysis, all scale attention models are implemented with the max pooling-based down-sampling method.

As we discussed in Subsection \ref{sec-III-I}, with the different definitions of attention, the 3D attention model can be generalized to pseudo 3D (channel-wise), 2D (spatial), and 1D (temporal) attention models with the decrease of model complexity. The results are shown in Table \ref{tab3_2}. They imply that applying different attention maps for each feature channel is better than applying the same attention map to all feature channels. In the case of the channel-wise attention model CW\_SAM\_2.5D (implemented as a depth-wise convolution \cite{Chollet2017} ), there is a slight decrease in the performance compared with the cross-channel attention model (3D), and the decrease is large when model parameters are shared in CW\_SAM\_2.5D (shared). Moreover, comparing results with CW\_SAM\_2.5D, SAM\_2D, and TAM\_1D, we can conclude that the channel and spatial factors should be taken into consideration in attention weight estimation.
\subsection{Effect of orthogonal regularization in attention map estimation}
\label{sub_orthogonal}
As presented in Subsection \ref{orthogonal}, in the estimation of attention maps for multiple channels, we added an orthogonal regularization to focus on different discriminative features in learning. With the variation of the regularization parameter $\lambda _2$, we carried out experiments to examine its effect. The results are shown in Table \ref{tab4}.
\begin{table}[tb]
\centering
\caption{Performance of cross channel 3D attention model with orthogonal regularization of attention maps (classification accuracy (standard deviation)) (\%)}
\begin{tabular}{|c||c|c|}
\hline
 $\lambda _2$ &Valid&Test\\
\hline
\hline
0.0001&81.9 (5.86) &78.1 (5.10)\\
0.001 &81.7 (5.75)&78.9 (6.10)\\
0.01&81.6 (5.70)&79.0 (5.49)\\
0.1&81.7 (5.19)&\textbf{79.4} (6.29)\\
1&81.4 (4.86)&79.0 (5.26)\\
5&81.0 (5.82)&78.4 (5.65)\\
\hline
\end{tabular}
\label{tab4}
\end{table}
The results show that adding orthogonal regularization for cross-channel attention map estimation is important for improving the performance. Moreover, setting $\lambda _2=0.1$ can obtain the best performance among the tried regularization parameters. These results also confirmed that constraining attention maps to focus on different features (channels) can help in improving discriminative feature extraction.
\subsection{Model complexity}
To check the complexity of different models, we showed the parameter size of each model in Table \ref{tabparam}.
\begin{table}[tb]
\centering
\caption{Numbers of model parameters (M: millions)}
\begin{tabular}{|c||c|}
\hline
Models & number of parameters \\
\hline
\hline
DCRNN (3 CNN blocks) & 1.484 M \\
\hline
DCRNN (4 CNN blocks) & 2.075 M \\
\hline
ResCNN & 0.510 M \\
\hline
CC\_SAM\_3D & 1.692 M \\
\hline
CW\_SAM\_2.5D & 0.518 M \\

\hline
CW\_SAM\_2.5D (shared) & 0.513 M \\
\hline
SAM\_2D &0.515 M \\
\hline
TAM\_1D & 0.512 M  \\
\hline
\end{tabular}
\label{tabparam}
\end{table}
The results in Tables \ref{tab1} and \ref{tab2} with reference to the number of the model parameter size show that skip connections for propagating cross-scale features to the top layers is important. Increasing model capacity does not always increase model performance. Furthermore, the performance improves consistently across different types of CSA models. This finding confirms that the performance improvement is attributed to the good model structure with the CSA models.
\subsection{Experiments on the DCASE2017\_T4 corpus}
We carried out experiments on the DCASE2017\_T4 data set and focused on the acoustic event tagging task (task A). The task is a large-scale sound event detection dataset, which consists of weakly labeled audio recordings. The audio data were excerpted from the YouTube video with transportation and warnings acoustic events. The design is to simulate the acoustic events that might be encountered in self-driving cars, smart cities, and related areas. All of the acoustic events were weakly labeled, and the timestamps were not provided. In contrast to the experiments on UrburnSound8K corpus, the recall, precision, and F1 values are used as evaluation metrics, which are defined in \cite{DCASE2017}, since the exact timestamps for each events are not provided.

On the basis of the analysis of the experimental settings for the UrburnSound8K corpus and with reference to the models proposed in \cite{XuNo117}  (top 1 model on task A of the DCASE2017\_T4 data set), we made a slight modification on the framework pipeline, as shown in Fig. \ref{figall}. Two BGRU layers were applied instead of one BGRU layer as used in the UrburnSound8K corpus. In addition, in the implementation of the ResCNN, we tried models with several residual blocks. The results are shown in Table \ref{tab_dcase2017}, where the ResCNN with the two-layered BGRU was termed ResCNN for simplicity. The CSA model with the ResCNN is presented as CC\_SAM\_3D. From the table, we can note that propagating cross-scale features with skip connections is important in discriminative feature extraction. The proposed CSA model with the ResCNN as a backbone further improved the performance and obtained state-of-the-art results in terms of recall and F1 scores.
\begin{table}[tb]
\centering
\caption{Performance of the comparative models and CSA models on DCASE2017\_T4 set)}
\begin{tabular}{|c||c|c|c|}
\hline
 Dev-set &Precision&Recall&F1\\
\hline
\hline
DCASE2017 Baseline \cite{DCASE2017}&7.8 &17.5&10.9\\
Gated\_CRNN\_LogMel \cite{XuNo117} &53.8 &60.1&56.7\\
Gated\_CRNN \_MFCC \cite{XuNo117}&51.7&52.5&52.1\\
\hline
ResCNN (2 Blocks) &52.0&56.4&54.1\\
ResCNN (3 Blocks) &54.2&57.3&55.7\\
ResCNN (4 Blocks) &53.7&\textbf{60.2}&56.8\\
\hline
CC\_SAM\_3D (3 Blocks) &\textbf{56.7}&60.1&\textbf{58.3}\\
\hline
\hline
Eval-set &Precision&Recall&F1\\
\hline
\hline
DCASE2017 Baseline \cite{DCASE2017}&15.0 &23.1&18.2\\
Gated\_CRNN\_LogMel \cite{XuNo117} &\textbf{58.9}&50.2&54.2\\
Gated\_CRNN \_MFCC \cite{XuNo117}&51.7&52.5&52.1\\
\hline
ResCNN (2 Blocks) &54.1&58.3&56.1\\
ResCNN (3 Blocks) &58.0&60.7&59.3\\
ResCNN (4 Blocks) &56.2&62.1&59.0\\
\hline
CC\_SAM\_3D (3 Blocks) &58.1&\textbf{62.9}&\textbf{60.4}\\
\hline
\end{tabular}
\label{tab_dcase2017}
\end{table}

\section{Discussion and conclusion}
\label{sec-V}
The CNN has been widely used as a feature processing block for robust and invariant feature extraction in AEC. With a preliminary analysis of the DCNN-based model for AEC, we showed that conventional classification models are actually based on features from a certain large-scale space where representations from such a scale space are suitable for improving the discrimination of most acoustic patterns with the cost of degradation for some other acoustic patterns. Nonetheless, cross-scale discriminative features should be explored from a wide range of scales. Although a direct link to multi-scale audio process is based on multi-scale time–frequency resolutions as used in \cite{ZhuZ2016,ZhuB2018}, the concept of cross-scale used in our study is different. Here, the cross-scale refers to specifying different weights on cross-scale features processed in a DCNN model. Different from the conventional usage of the DCNN-based model for AEC, we proposed a scale attention model for an efficient propagation of cross-scale features from the bottom to top layers. With an attention weighting of features in consecutive layers, we can propagate discriminative features from different scales to the final feature representation. Through formulations (Eqs. (\ref{eq4}), (\ref{eq5}), (\ref{eq6}), and (\ref{eq7})), we derived that the proposed CSA can be regarded as an extension of the conventional ResCNN. In addition, in the calculation of the attention weights, either temporal or spatial context information can be combined to estimate the attention weights. From this point of view, the proposed CSA model can be regarded as a generalized attention model of either the temporal or spatial attention model.

We carried out experiments to test the proposed framework based on the CSA model and confirmed that adaptively incorporating cross-scale features with attention progressively performs better than only using a fixed-scale propagation in feature extraction (as used in ResCNN). Moreover, our experiments show that explicitly incorporating spatial attention performs better than incorporating only temporal attention and weighting differently for each feature channel performs better than using the same attention to all feature channels. However, incorporating more feature information with a large increase in the model parameter size sometimes does not necessarily improve the performance. How to explicitly incorporate much more information with a slight increase in the model parameter size is one of our future works. Moreover, we will further investigate the effective factors involved in the design of the CSA blocks, for example, how to design the residual branch as shown in panel (b) of Fig. \ref{fig_resnet} and how to include context information either with local or global attention.

\end{document}